

\font\rmB=cmr9 scaled\magstep4

\magnification=1200
\parskip=5pt plus 2pt
\vsize=9.1 true in
\hsize=6.2 true in
\parindent=20pt
\baselineskip=17pt


\newcount\notenumber
\def\resetnotenumber{\notenumber=1}
\def\note#1{{\baselineskip=12pt\footnote{$^{\the\notenumber}$}{#1}}
\advance\notenumber by1}
\resetnotenumber
%


\def\R{{\rm I\! R}}
\def\\C{{\rm C}}
\def\C{\mkern1mu\raise2.2pt\hbox{$\scriptscriptstyle|$}{\mkern-7mu\rm C}}
\def\dalemb #1#2 {{vbox {\hrule heigh.#2pt
	\hbox {vrule width.#2pt heigh #1pt \kern #1pt
	\vrule width. #2pt} \hrule height.#2pt}}}
\def\Z{{\rm \bf Z}}
\def\pa{\partial}

\def\and{{\rm\ and\ }}

\def\abs #1{\vert {#1}\vert}
\def\frac#1#2{{#1 \over #2}}
\def\half{{\frac 12}}
\def\twid #1{\tilde {#1}}

\def\what #1{\widehat {#1}}
\def\pde #1#2{\partial_{#2} #1}


\def\a {\alpha}
\def\b {\beta}
\def\g {\gamma}

\def\om {\omega}
\def\d {\delta}
\def\e {\epsilon}

\def\l {\lambda}

\def\s {\sigma}
\def\t {\tau}
\def\vf {\chi}

%
%

\def\Si {\Sigma}

\def\Om {\Omega}

\def\pi {\twid p}

%
%

\def\am {\cal A}

\def\em {\cal E}

\def\On {1\over \Om}

%
%
\rightline {Imperial/TP/91-92/7}
\rightline {QMW/PH/92/7}
\rightline {April 1992}
\vglue 1in
\centerline {\rmB Ashtekar Formulation of 2+1 Gravity}
\centerline {\rmB on a Torus}
\vskip 0.5in
\centerline {Nenad Manojlovi\'c}
\centerline {\it The Blackett Laboratory, Imperial College,}
\centerline {\it Prince Consort Road, London SW7 2BZ, U. K.}
\vskip 0.2in
\centerline {and}
\vskip 0.2in
\centerline {Aleksandar Mikovi\'c}
\centerline {\it Department of Physics, Queen Mary and Westfield College,}
\centerline {\it Mile End Road, London E1 4NS, U. K.}
\vskip 0.5in
\centerline {\bf Abstract}
\vskip 0.01in
Pure (2+1)-dimensional Einstein
gravity is analysed in the Ashtekar formulation, when the spatial manifold
is a torus. We have found a set of globally defined
observables, forming a closed algebra. This allowed us to solve the
quantum constraints, and to show that the reduced phase space of the Ashtekar
formulation is greater then the corresponding space of the Witten formulation.
Furthermore, we have found a globally defined time variable which satisfies all
the requiriments of an extrinsic time variable in quantum gravity.
\medskip
\smallskip
\smallskip
\smallskip
\phantom{.}
\eject

\noindent{\bf 1. Introduction}

$(2+1)$-dimensional Einstein
gravity [1] is an example of a toy model useful
for studying some of the conceptual problems of $(3+1)$-dimensional
quantum  gravity. The (2+1)-dimensional theory
is a tractable model, since it has
finitely many physical degrees of freedom, while on the other hand
it retains some basic features of the full theory.

The Ashtekar formulation of $3+1$ Einstein gravity has brought
significant simplifications of the constraints [2], and a large class
of solutions of all the quantum
constraints has been found [3]. However,
since the basic object in the theory is a connection, it is not clear how
to interpret the results in terms of more familiar objects like metrics.
Therefore a study of $2+1$ gravity in the Ashtekar formulation
may give some insights into these problems.

The connection formulation of $2+1$ gravity was first analysed
by Witten [4], who also showed that
the theory can be written as a Chern-Simons gauge theory, based on the
three-dimensional Poincare group $ISO(1,2)$.
The Witten formulation was inspired by the
Ashtekar formulation of $3+1$ gravity, but it was Bengtsson who
showed that the theory can be formulated in terms of the $(2+1)$-dimensional
analogs of the
Ashtekar constraints [5]. He also showed that the two formulations
are equivalent for
the non-degenerate spatial metrics, and he argued that Witten's formulation
is more restrictive than the Ashtekar formulation in the
case of the degenerate metrics. Although
the reduced phase space and quantization of the Witten theory was studied by
many authors [4,6,7,8], this cannot be said for the Ashtekar theory.

In this paper we study the reduced phase space and quantization of the Ashtekar
formulation of $2+1$ gravity in the case when the spatial manifold is a
torus. The advantage of the torus case is that one can explicitely write
down the global variables which can be used to define the
global physical degrees of freedom. Our choice of the global variables is
motivated by the variables used in the Ashtekar formulation of Bianchi
cosmologies in $3+1$ dimensions [9]. Consequently, the theory becomes
equivalent to a pair of particles propagating in a three-dimensional
Minkovski space, subjected to constraints. By constructing an algebra of
observables we solve the quantum theory and find the reduced phase space (RPS),
and consequently show that the RPS of the Ashtekar theory is larger than the
RPS of the Witten theory. Furthermore, we construct a time variable out of
the elements of the algebra of observables, and write down the corresponding
Schrodinger equation, which is equivalent to a $(2+1)$-dimensional massless
Klein-Gordon equation.

In section 2 we discuss the canonical formalism for the Palatini form of
the Einstein-Hilbert action. We derive the Ashtekar form of the constraints
from the Palatini action by a judicious choice of the normal vector of the
foliation. In section 3 we specialize to the case when the spatial manifold
is a torus, and introduce global variables by using the cohomology basis forms.
We then derive the corresponding
effective finite-dimensional theories describing the Witten and
the Ashtekar formulations, and then rederive the reduced configuration space
for the Witten theory. In section 4 we study the Ashtekar constraints
in terms of our variables. We use the Dirac method to solve the gauge
constraint, and from the form of the solution we deduce the
gauge invariant coordinates. Then we solve the
scalar constraint by constructing an algebra of observables, which turns out
to be isomorphic to the three-dimensional Poincare algebra. The scalar
constraint then becomes the masslessness condition. This gives only two
inequivalent irreducible representations, a scalar and a spinor, which
correspond to two inequivalent quantizations of the theory.
The corresponding massless Klein-Gordon equation
can be then used to define a time variable. In section five we present our
conclussions.

\noindent{\bf 2. Canonical Formalisam}

We start from the Palatini action
$$
\eqalignno {
S ( e , \om ) &= {\half} \int _{\cal M} d ^3 x \, \e ^{\l \mu \nu} e_{\l a}
F_{\mu \nu}^{a} 	\, \, \, ,				&(2.1) \cr}
$$
where $\l , \mu , \nu = 0, 1, 2$ are the coordinate indices on a three
dimensional space-time manifold ${\cal M}$, while $a , b , c = (0) , (1) ,
(2)$ are the Lie algebra $L(SO(2,1))$ indices. $\eta$ is the
Minkowski metric on $L(SO(2,1))$ and $\e ^{\l \mu \nu}$
is the metric independent totally anti-symmetric tensor density of
weight one, with $\e _{012} = 1$. ${e _{\l}} ^a$ is a triad, a
one-form on ${\cal M}$ which takes
its values in the Lie algebra $L(SO(2,1))$. $F _{\mu \nu} ^a$ is the
curvature for a spin connection ${\om _{\mu}} ^a$ on
${\cal M}$. In order to obtain the canonical formulation of the theory
we have to assume
that ${\cal M}$ is topologicaly equivalent to $\Si \times \R$,
where $\Si$ is a two-dimensional manifold. We take $\Si$ to be a compact
manifold. Let $X$ be
a diffeomorphism $X: \Si {\times} \R \to {\cal M}$, such that it defines a
foliation.
We then choose a Lorentzian metric $g _{\mu \nu}$ on ${\cal M}$ such that
each leaf $\Si _t \equiv X _t ( \Si )$ is a space-like submanifold of
${\cal M}$ with respect to the metric $g _{\mu \nu}$. Next we introduce
the deformation vector of the foliation
$t ^{\mu} \equiv {\dot X} ^{\mu}$,
and decompose it into two components, one of which lies along
the surface $\Si _t$ and the other is normal to $\Si_t$
$$\eqalignno {
t ^{\mu} \equiv {\dot X} ^{\mu} &= N g ^{\mu \nu} n _{\nu}
+ N ^i \pde{X ^{\mu}}i        \, \, \, .&(2.2) \cr}$$
$i=1,2$ is the coordinate index on $\Si$,
$N$ is the lapse function, $N^i$ is the shift vector
and $n _t$ is the normal to the foliation, defined
to be the unique one-form $n _t$ on ${\cal M}$
which satisfies the equations
$$
\eqalignno {
n _{\mu}\pde{X ^{\mu}}i &= 0		\, \, \, ,		&(2.3a) \cr
g ^{\mu \nu} n _{\mu} n _{\nu} &= -1	\, \, \, .		&(2.3b) \cr}
$$
Equation $(2.3a)$ means that $n _t (x)$ is normal to
the hypersurface $\Si _t$, while the equation $(2.3b)$ is the normalization
condition for $n _t$. The induced metric on $\Si$ is given by
$$g _{ij} = \pde{X ^{\mu}}i \pde{X ^{\nu}}j g _{\mu \nu}\eqno(2.4)$$
and the spatial components of $\om$ and $e$ are defined as
$${A _i} ^a = \pde{X ^{\mu}}i {{\om} _{\mu}} ^a \quad,\quad
e _{ia} = \pde{X ^{\mu}}i e _{\mu a}\quad.\eqno(2.5)$$

The action (2.1) can be now rewritten as
$$S = \int dt \int _{\Si} d^2 x \, \Bigl ( {E ^i} _a {{\dot A} _i} ^a
- Nn^{\mu}\om_{\mu} ^a G _a
- N ^i (\half e_i^a \e^{jk} F_{jk}^a - A_i^a G_a )
- N n^{\mu} e_{\mu}^a \e^{ij} F_{ij}^a \Bigr ), \eqno(2.6)$$
where $\e^{ij}$ is the Levi-Civita tensor density on $\Si$,
${E ^i} _a = \e ^{ij} e _{ja}$ and
$$G _a = \pde {{E ^i} _a} i + {\e _{ab}} ^c  {A _i} ^b {E ^i} _c \eqno(2.7)$$
are the generators of the local Lorentz transformations.
By taking $X^{\mu}$ to be the trivial foliation $X^0 =t,X^i = x^i$, one
recovers from (2.6) the Witten formulation of $2+1$ gravity [4], since then
$N^i=0$ and the constraints are
$$\eqalignno {
G _a &= \pde {{E ^i} _a} i + {\e _{ab}} ^c  {A _i} ^b {E ^i} _c \approx 0
\, \, \, ,	&(2.8a) \cr
C ^a &= \e ^{ij} F _{ij}^a \approx 0	\, \, \, . 	&(2.8b) \cr}$$
However, one does not have to choose the
foliation $X^{\mu}$ explicitely.
By choosing the normal $n$ to be
$$\eqalignno { n _{\mu} &= - {\half} {\abs g} ^{-\half} \e ^{abc} \e ^{jk}
e _{\mu a} e _{jb} e _{kc}\, \, \, ,&(2.9) \cr}$$
where ${\abs g}$ is the determinant of the spatial metric $g_{ij}$
and ${\e _{ab}} ^c$ are
the structure constants of the Lie algebra $L(SO(2,1))$, satisfying
$$
\eqalignno {
{\e _{ab}} ^c {\e _{cd}} ^e &= - ( \eta _{ad} \d ^e _b - \eta _{bd}
\d ^e _a )	 \, \, \, , 					&(2.10) \cr}
$$
the equations (2.3) will be satisfied identicaly. The action (2.1)
then takes the following form
$$
\eqalignno {
S &= \int dt \int _{\Si} d^2 x \, \Bigl ( {E ^i} _a {{\dot A} _i} ^a
- N ^a G _a - N ^i G _i - N G _0 \Bigr )	\, \, \, ,	&(2.11) \cr}
$$
where $N^a = Nn^{\mu}\om_{\mu} ^a$ and
$$
\eqalignno {
G _a &= \pde {{E ^i} _a} i + {\e _{ab}} ^c  {A _i} ^b {E ^i} _c \approx 0
\, \, \, ,							&(2.12a) \cr
G _i &= F _{ij} ^a {E ^j} _a - {A _i} ^a G _a \approx 0
\, \, \, ,							&(2.12b) \cr
G _0 &= {\e _a} ^{bc} F _{ij} ^a {E ^i} _b {E ^j} _c \approx 0
\, \, \, .							&(2.12c) \cr}
$$
The $(E,A)$ variables together with the
constraints $(2.12)$ are the three-dimensional
analogs of the Ashtekar formulation of
general relativity [5]. An important difference between the $3+1$ and the $2+1$
case is that in the $2+1$ case the $(E,A)$ variables are real.
Note that the Ashtekar constraints (2.12) are
equivalent to the Witten constraints (2.8) if $|g|\ne 0$, since (2.12b) and
(2.12c) imply (2.8b) if $det(E^i_a E^{ja})\ne 0$ [5]. This can be also seen
from the equation (2.9), which is defined for $|g|\ne 0$, since classically any
two foliations give equivalent theories.
However, in the
region of the phase space where $|g|=0$, these theories are different, as the
subsequent analysis will show.

Pure three-dimensional Einstein
gravity has no propagating degrees of freedom.
In the covariant approach this fact follows from the identity
${R ^{\mu \nu}} _{\a \b} = \e ^{\mu \nu \s} \e _{\a \b \l}
{G ^{\l}} _{\s}$, where ${R ^{\mu \nu}} _{\a \b}$ is the curvature
tensor and $G ^{\l}_{\s}$ is the Einstein tensor,
so that the Einstein equations in empty space imply vanishing of
the curvature tensor [1]. In the canonical approach, this is a consequence
of the constraints (2.8) or (2.12). Namely, there are six constraints per
space point and there is the
same number of the configuration space variables. This means
that there are no local physical degrees of freedom. However, certain
global degrees of freedom survive, which make the theory non-trivial.

\noindent{\bf 3. Classical Theory on a Torus}

In order to see how the global physical degrees of freedom arise,
it is instructive to study the case when $\Si$ is a torus.
Torus is a globally homogenous space, and therefore one can find
globally defined one-forms $\vf ^{\a}(x), \a = 1 , 2$ which satisfy the
Maurer-Cartan equations
$$\eqalignno {
d {\vf} ^{\a} &= 0 \, \, . &(3.1) \cr}$$
The corresponding vector fields $L _{\a}(x) , \a = 1 , 2$, satisfy
$$
L^i_\a \vf_i^\b = \d_\a^\b \quad,\quad
[ L _{\a} , L _{\b} ] = 0 	\, \, \, . \eqno(3.2)
$$
A gauge connection on the torus can be always written as a linear
combination of the $\vf$'s, with coordinate independent coeficients,
up to a gauge transformation, since there is enough gauge symmetry in
the theory. Consequently we can write
$$
\eqalignno {
{A _i} ^a ( x , t ) &= {{\am } ^a}_\a (t) \,   {{\vf} _i}^\a (x)
\, \, \, ,							&(3.3a) \cr
{E^i}_a ( x , t ) &= \abs {\vf} \, {{\em} _a}^\a (t) {L ^i} _\a (x)
\, \, \, ,							&(3.3b) \cr}
$$
where
$$
\eqalignno {
{{\am} ^a} _{\a} &= {\On} \int _{\Si} d^2 x \, \abs {\vf} \,
{A _i} ^a {L ^i} _{\a}		\, \, \, ,			&(3.4a) \cr
{{\em} _a} ^{\a} &= {\On} \int _{\Si} d^2 x \, {E ^i} _a {\vf _i} ^{\a}
				\, \, \, ,			&(3.4b) \cr}
$$
and $\Om = \int _{\Si} d ^2 x   | {\vf} |$. The $(\am,\em)$ variables
are global and coordinate indipendent, and can be used to define the
global physical degrees of freedom. Note that in the case
of $(3+1)$ gravity the choice
(3.3) is a genuine reduction of the degrees of freedom [10], while in the
$2+1$ case (3.3) can be thought of as a partial gauge choice.

By inserting the expressions (3.3) into the action (2.11),
we obtain the effective finite-dimensional theory, defined by
the action
$$
\eqalignno {
{\it S}  &= \int _{t_1} ^{t_2} \! dt \, \Bigl ( \Om \,
{{\em} _a} ^{\a} {{\dot {\am}} ^a} _{\a} - m ^a {\cal G} _a - n {\cal S} \Bigr
)
\, \, \, ,&(3.5) \cr} $$
where the Lagrange multipliers $\{ m ^a , n \}$ impose the first class
constraints
$$
\eqalignno {
{\cal G} _a &= {\e _{ab}} ^c {{\am} ^b} _{\a} {{\em} _c} ^{\a}
\, \, \, ,&(3.6a) \cr
{\cal S} &= \bigl ({\am}^a_\a {\em}_a^\a \bigr ) ^2 -
{\am}^a_\a {\em}^\a_b {\am}^b_\b {\em}^\b_a
\, \, \, . &(3.6b) \cr}$$
The constraint (3.6a) is the global remnant of the gauge constraint $G_a$,
while (3.6b) is the global remnant of the scalar constraint $G_0$. The global
remnant of the diffeomorphism constraint $G_i$ is proportional to ${\cal G}$
and therefore not indipendent.
The constraints $(3.6)$ are irreducible and first class, and form the
following Poisson bracket algebra
$$
\eqalignno {
\{ {\cal G} _a , {\cal G} _b \} &= {\e _{ab}} ^c \, {\cal G} _c
\, \, \, ,							&(3.7a) \cr
\{ {\cal G} _a , {\cal S} \} &= 0	\, \, \, ,		&(3.7b) \cr}
$$
which can be derived by using the basic canonical brackets
$$ \{ {\am}^{a}_{\a} , {\em}_{b}^{\b} \} = \d_{\a}^{\b} \d_{b}^{a} \quad.
\eqno(3.8)$$

There are four constraints and six degrees of freedom in the action (3.5),
so that leaves only
two physical degrees of freedom. One can obtain the same result in the
Witten formulation. By inserting the expressions (3.3) into the Witten
constraints (2.8), one gets the action
$$\eqalignno { {\it S}  &= \int _{t_1} ^{t_2} \! dt \, \Bigl ( \Om \,
{{\em} _a} ^{\a} {{\dot {\am}} ^a} _{\a} - m ^a {\cal G} _a
- n _a {\cal C} ^a \Bigr )\, \, \, ,	&(3.9) \cr}$$
where
$$
\eqalignno {
{\cal G} _a &= {\e _{ab}} ^c {{\am} ^b} _{\a} {{\em} _c} ^{\a}
\approx 0		\, \, \, ,				&(3.10a) \cr
{\cal C} ^a &= \e ^{\a \b} \, {\e ^a} _{bc} {{\am} ^b} _{\a} {{\am} ^c} _{\b}
\approx 0		\, \, \, .				&(3.10b) \cr}
$$
The constraints $(3.10)$ are first class and their Poissson bracket
algebra is
$$
\eqalignno {
\{ {\cal G} _a , {\cal G} _b \} &= {\e _{ab}} ^c \, {\cal G} _c
\, \, \, ,							&(3.11a) \cr
\{ {\cal C} ^a , {\cal G} _b \} &= {\e ^a} _{bc} \, {\cal C} ^c
\, \, \, .							&(3.11b) \cr}
$$
However, the constraints (3.10) are not independent, since they satisfy
$$ {\am}_{\a}^{a} G_{a} - {\em}^{\a}_{a} {\cal C}^{a} = 0 \quad. \eqno(3.12)$$
This means that there are $3+3 -2 =4$ indipendent constraints, which gives
$6-4 =2$ physical degrees of freedom. This agrees with the result obtained
in [8], where the reduced phase space was analysed through the Wilson loop
variables (trace of the holonomies), as well as with the dimension of the space
of flat gauge-inequivalent connections on the torus.

In order to show that
the reduced configuration space defined by the constraints (3.10) is the same
as the moduli space of flat connections on a torus, one can use the
same method as the one employed in [8], i.e. analyse the holonomy variables
$$ {\rm Tr}\, U_\g (A) = {\rm Tr}\,{\rm P}\,{\rm exp} \int_\g A_i^a \t_a dx^i
\quad,\eqno(3.13)$$
where $\t_a$ are matrices representing the $LSO(1,2)$ algebra, and $\g$ is a
loop on the torus. By chosing a homology basis of loops $\{\g_\a \}$ such that
$$ \int_{\g_\a} \chi^\b = \d_\a^\b \quad,\eqno(3.14)$$
and by using (3.3a), we get
$$ T_\a \equiv {\rm Tr}\, U_{\a} = {\rm Tr}\,{\rm exp} ( {\am}_{\a}^{a} {\t}_a
) \quad.\eqno(3.15)$$
$T_\a$ have vanishing Poisson brackets with the constraints (3.10) and
therefore represent observables in the Witten formulation.
A connection $A={\am}_{\a}^{a} \t_{a} {\vf}^{\a} = \theta_{\a}\vf^\a$
is flat if $\theta_\a$ belongs to an Abelian subalgebra of $LSO(1,2)$.
Consequently, there will be three sectors, depending on whether
$\vec{{\am}_\a}$
is space-like, time-like or null. In terms of the $U_{\a}$ these sectors
correspond to $SO(2),SO(1,1)$ and $GL(1)$ subgroups, respectively.
This gives for the reduced configuration
space
$$ C^*_W = \Bigl( S^1\times S^1 \Bigr)
\cup \Bigl( \R_+ \times \R_+ \Bigr) \cup \R \quad.
\eqno(3.16)$$

In order to obtain the reduced configuration space $C^*_A$ for the theory
defined by the Ashtekar constraints (3.6), we will also use the method of
finding the algebra of observables.

\noindent{\bf 4. Dirac Analysis}

Let us
introduce a notation ${\vec {\am}}_{\a} \equiv {\vec r} _{\a}$ and
${\vec {\em}} ^{\a} \equiv {\vec p} _{\a}$,
so that the system
defined by the action $(3.5)$ is equivalent to a system of two particles
in the three dimensional Minkowski space, with the constraints
$$
\eqalignno {
{\vec {\cal G}} &= \sum _{\a =1} ^2 {\vec r} _{\a} \times {\vec p} _{\a}
\approx 0 	\, \, \, ,					&(4.1a) \cr
{\cal S} &= \sum _{\a , \b =1} ^2 \Bigl( {\vec r} _{\a} \times
{\vec r} _{\b} \Bigr ) \cdot \Bigl ( {\vec p} _{\a} \times
{\vec p} _{\b} \Bigr )
= 2 \Bigl( {\vec r} _{1} \times
{\vec r} _{2} \Bigr ) \cdot \Bigl ( {\vec p} _{1} \times
{\vec p} _{2} \Bigr )
\approx 0 	\, \, \, .		&(4.1b) \cr}
$$
The constraint $(4.1a)$ can be interpreted as the total angular momentum, and
it generates the $SO(1,2)$ rotations.
The quantum theory can be defined by promoting $(r,p)$ variables into operators
satisfying the Heisenberg algebra
$$ [ x_{\a}^a , p_{\b}^b ] = i \d_{\a\b} \eta^{ab} \quad. \eqno(4.2)$$
Next we choose the normal ordering for
the constraints (4.1) such that
$$
\eqalignno {
{\what {\vec {\cal G}}} &= \sum _{\a =1} ^2 {\what {\vec r}} _{\a}
\times {\what {\vec p}} _{\a}
\, \, \, ,&(4.3a) \cr
{\what {\cal S}} &= {\what x} _1 ^{[a|} {\what x} _2 ^{|b]}
{\what p} _{1a} {\what p} _{2b}
\, \, \, .			&(4.3b) \cr}$$
The constraint algebra is preserved under this ordering. As a next
step in the Dirac procedure,
we impose the constrains $(4.3)$ on the physical
states in the coordinate representation. The gauge constraint gives
$$
\eqalignno {
i{\what {\vec {\cal G}}} \psi ({\vec r} _1 , {\vec r} _2) &=
\sum _{{\a} =1} ^2 {\vec r} _{\a} \times {\partial \psi \over
\partial {\vec r} _{\a}} = 0		\, \, \, .		&(4.4) \cr}
$$
The general solution of the system $(4.4)$ is
$$\eqalignno {
\psi &= \psi ( a _1 , a _2 , a _3 ) &(4.5) \cr}$$
where $a _1 = {\vec r}_1 \cdot {\vec r}_1 , a _2 ={\vec r}_2\cdot{\vec r}_2$
and $a _3 = {\vec r} _1\cdot {\vec r} _2 $. The $a_i$ variables can be defined
classically, and they represent an irreducible set of rotationally invariant
observables. Since $a_i$ form an Abelian algebra, they can be used as
coordinates on the reduced phase space obtained by solving the rotational
constraint (4.1a). The scalar constraint (4.1b) is rotationally
invariant, and therefore can be expressed as a function of $a_i$ and their
canonically conjugate momenta. This implies that the action of the scalar
constraint operator on the states (4.5) will be a function of $a_i$ an their
derivatives, which can be checked directly
$$
\eqalignno {
{\what {\cal S}} \psi ( a _1 , a _2 , a _3 ) &=
\bigl ( a _1 a _2 - (a _3) ^2 \bigr )
\, \bigl ( 4 {\partial ^2 \psi \over \partial a _1 \partial a _2} -
{\partial ^2 \psi \over (\partial a _3) ^2} \bigr ) = 0
\, \, \, .							&(4.6) \cr}
$$

It is convenient to perform a coordinate transformation
$( a _1 , a _2 , a _3 ) \to ( {\l} _0 , {\l} _1 , {\l} _2 )$ where
$$
\eqalignno {
{\l} _0 &= {\half} ( a _1 + a _2 ) 	\, \, \, ,		&(4.7a) \cr
{\l} _1 &= {\half} ( a _1 - a _2 )	\, \, \, ,		&(4.7b) \cr
{\l} _2 &= a _3				\, \, \, .		&(4.7c) \cr}
$$
Then (4.6) can be rewritten as
$$
\eqalignno {
{\what {\cal S}} \psi ( {\l} _0 , {\l} _1 , {\l} _2 ) &=
\bigl ( ({\l} _0) ^2 - ({\l} _1) ^2 - ({\l} _2) ^2 \bigr ) \,
\bigl ( \partial ^2 _0 \psi - \partial ^2 _1 \psi - \partial ^2 _2 \psi _2
\bigr ) = 0			\, \, \, .			&(4.8) \cr}
$$
Note that the transformation (4.7) defines a canonical transformation on the
reduced phase space $(a, P_a ) \to (\l , P_{\l})$ such that
${\cal S} = \l^2 P_{\l}^2 $.

In order to find the algebra of observables, note that (4.8) is almost a
massless Klein-Gordon equation (KGE).
Besides the Poincare symmetry, the massless KGE has a larger group of
symmetries, generated by the conformal group.
Let us consider the generators ${\what \Pi} _i , {\what L} _i, {\what K} _i ,
{\what \Delta}$ of the conformal algebra, which in our case is
the $L(SO(3,2))$ Lie algebra, defined by
$$
\eqalignno {
i{\what \Pi} _i &\equiv \partial _i
\, \, \, ,							&(4.9a) \cr
i{\what L} _i &\equiv {\e _i} ^{jk} \l _j \partial _k
\, \, \, ,							&(4.9b) \cr
i{\what K} _i &\equiv - \half \l ^2 \partial _i + \l _i \l ^j \partial _j
+ d \, \l _i    		\, \, \, ,			&(4.9c) \cr
i{\what \Delta} &\equiv \l ^i \partial _i + d	\, \, \, ,	&(4.9d) \cr}
$$
where $d$ is a normal ordering constant [11,12].
${\what L} _i$ are the Lorentz generators, ${\what K} _i$ are
the conformal boosts and ${\what \Delta}$ is the dilaton generator.
A straightforward calculation shows that ${\what L} _i$ and ${\what \Delta}$
commute with the scalar constraint ${\what {\cal S}}$
$$
\eqalignno {
[{\what {\cal S}} , {\what L} _i ] &= 0		\, \, \, ,	&(4.10a) \cr
[ {\what {\cal S}} , {\what \Delta} ] &= 0	\, \, \, .	&(4.10b) \cr}
$$
The commutator $[{\what {\cal S}},K_i ]$
is not equal to zero identically
$$
\eqalignno {
[ {\what {\cal S}} , {\what K} _i ] &\approx {2 \l ^2 \over i} ( d -
{\half} ) \partial _i 				\, \, \, , 	&(4.11) \cr}
$$
where $\approx$ means equality up to terms proportional to ${\what {\cal S}}$.
However, if we set $d = \half$ than the right hand side of $(4.10)$
is weakly equal to zero. Therefore
${\what L} _i, {\what K} _i$ and ${\what \Delta}$ are observables for $2+1$
gravity on a torus, and they form a closed subalgebra of the conformal
algebra
$$
\eqalignno {
[{\what L} _i, {\what L} _j] &= i{\e _{ij}} ^k {\what L} _k
\, \, \, ,							&(4.12a) \cr
[{\what L} _i, {\what K} _j] &= i{\e _{ij}} ^k {\what K} _k
\, \, \, ,							&(4.12b) \cr
[{\what K} _i, {\what K} _j] &= 0	\, \, \, ,		&(4.12c) \cr
[{\what L} _i , {\what \Delta}] &= 0	\, \, \, ,		&(4.12d) \cr
[{\what K} _i , {\what \Delta}] &= i {\what K} _i    \, \, \, ,	&(4.12e) \cr}
$$
which is known as the Weil algebra. Note that $K , L$ and $\Delta$ can be
defined as classical observables, by dropping the $i$ from the expressions
(4.9), and by replacing the $\partial_i$ with $P_{\l_i}$. The constant $d$
is then zero, while in the quantum case it
takes the value $\half$, in order for the Weil symmetry
not to be anomalous.

Note that the scalar constraint can be written as
$$
\eqalignno {
{\what {\cal S}} &= {\what L} ^2 - {\what \Delta} ^2
- {1\over 4}			\, \, \, ,			&(4.13) \cr}
$$
which means that the elements of the Weil algebra are not independent on
the constraint surface. Relation (4.13) implies that a smaller set of
observables is more appropriate, which is the Poincare subalgebra
$\{K_i ,L_i \}$.
Furthermore, the following relation is satisfied
$$ {\what K}^2 = -\frac14 \l^2 {\what {\cal S}} \quad,\eqno(4.14)$$
so that only the massless representations of the Poincare algebra are relevant.

In the quantum theory only the unitary irreducible representations
(UIR) are relevant, which for the case of the three-dimensional Poincare group
were classified in [13]. Since the relevant UIR are characterized by
$K ^2 = 0$ and $K_0 > 0 $,
we set $K _0 = + \sqrt { {K _1} ^2 + {K _2} ^2 }$.
There are only two inequivalent massless UIR, the scalar and the spinor UIR.
The wavefunctions belonging
the scalar UIR are scalars $\phi ( K _1 , K _2)$.
In this representation ${\what K} _i$ are multiplicative operators
and ${\what L} _i$ act as
$$
\eqalignno {
{\what L} _0 &= i ( K _1 \partial _2 - K _2 \partial _1 )
\, \, \, ,							&(4.15a) \cr
{\what L} _1 &= -i K _0 \partial _2		\, \, \, ,	&(4.15b) \cr
{\what L} _2 &= i K _0 \partial _1		\, \, \, ,	&(4.15c) \cr}
$$
where $\partial _1 = {\partial \over \partial K _1}$ and similarly for
$\partial _2$.
The wavefunctions that belong to the spinor UIR
are ``almost'' scalars in the sense that the infintesimal
operators ${\what L} _i$ are the same as in $(4.15)$. However, under a finite
rotation  the wavefunction $\psi( K _1 , K _2)$ may change its sign,
depending on the
value of the rotational parameter [13]. The reason for this pecuilar behaviour
of massless spinors in three
dimensions can be atributed to the fact that the little group
$SO(D-2)$ is not defined in $D=3$, but instead becomes a finite group $\Z_2$
[14].

Note that the physical state condition
$K^2=0$ becomes a Klein-Gordon equation
in the coordinate representation $K_i =i{\partial\over \partial X^i}$,
where $X^i$ are canonically conjugate variables to $K_i$.
Consequently $X^0$ can be interpreted as a time variable of $2+1$ gravity
on a torus.
It is not difficult to see that $X^0$ satisfies all the
requirements of an extrinsic time variable, according to Kuchar [15].
One can write a Schrodinger type equation governing the evolution
of the wavefunction of the toroidal universe
$$ \Bigr( i{\pa\over\pa T} - {\what H} \Bigl) \Psi ( T, X^\a ) = 0 \quad,
\eqno(4.16)$$
where $T=X^0$ and $H = K_0 = \sqrt{ K_1^2 + K_2^2 }$. ${\what H}$ is hermitian
with respect to the scalar product
$$ <\Psi | \Phi > = \int d^2 X \,\Psi^* (X , T) \Phi (X ,T)\quad,\eqno(4.17)$$
which is consequently indipendent of the time $T$.
The states $\Psi (X,T)$ can be expressed as linear combinations of the states
$\phi (K_1 , K_2)$ as
$$ \Psi (X,T) = \int d^2 K \, e^{i(K_0 T - K_\a X^\a )} \phi (K_\a ) \quad.
\eqno(4.18)$$
Note that in the standard treatment of the KG equation [16], there is an
additional factor of $K_0^{-\half}$ in the integrand of (4.18), which comes
from
the manifestly Poincare invariant form of (4.18). However, this factor is only
important for displaying the manifest Poincare invariance, and can be removed
by rescaling the $\phi (K)$. Then $X_\a$ can be represented as
$i{\pa\over \pa K_\a}$, which is hermitian with respect to the scalar product
$$ <\psi | \phi > = \int d^2 K \,\psi^* (K) \phi (K) \quad.\eqno(4.19)$$
The so called Newton-Wigner operators [16] correspond to representing
the $X_\a$ as
hermitian operators with respect to the scalar product
$$ <\psi |\phi > = \int d^2 K \, K_0^{-1}\psi^* (K) \phi (K)
\quad,\eqno(4.20)$$
but according to the Stone-von Neumann theorem any such two representations
are unitary equivalent.

\noindent{\bf 5. Conclusions}

Our analysis has shown that when $\Si$ is a torus, the reduced phase space
of the Ashtekar formulation of $2+1$ gravity
is not the same as the RPS
of the Witten formulation, something to be expected
on general grounds [5], but never demonstrated before. In the Ashtekar theory
the reduced configuration space is just $\R^2$, which is bigger but simpler
space than (3.15).

This result can be understood from the point of view of the
loop variables approach of [8].
Note that in [8] it was the Witten formulation of $2+1$ gravity which
was analysed. This simplifies the loop variables analysis, since then the
trace of the holonomy (3.13) is an observable, while in the Ashtekar
formulation (3.13) is not an observable (it transforms under the
diffeomorphisms). One can still use the same methods to analyse the theory,
but now the similarity with the $3+1$ case is even greater, and one can
use exactly the same reasoning as in the $3+1$ case.
Namely, in analogy with the $3+1$ case [3],
a solution of all the quantum
constraints is a wavefunction with a support on
the smooth link classes of the spatial manifold.
The analogs of the link classes for two-dimensional manifolds
are the homotopy classes. The homotopy classes of $\Si$
are directly related to the space of gauge inequivalent flat connections
on $\Si$ [8],
and therefore it seems that the Ashtekar formulation has the same reduced phase
space as the Witten formulation. However, it is known that in the $3+1$ case
the wavefunctions with supports on the
smooth link classes are not the most general solutions of the
quantum constraints (one can have wavefunctions with supports on the
generalized link classes, which include intersecting loops) [17].
This means that more general solutions also exist in the
$2+1$ case, which agrees with our result that $C^*_W \subset C^*_A $,
so that the wavefunctions with supports on the
smooth homotopy classes do not exhaust the whole space of
solutions for the Ashtekar constraints.

Note that $X^0$ represents a time variable in the Ashtekar formulation on a
torus. Obtaining the form of $X^0$ in terms of the loop variables
may give a hint about the time variable in the $3+1$
case. Another important point is that we were able to find
the usual quantum mechanical interpretation of an abstract scalar
product which had been obtained from the UIR of the algebra of observables.
However, in order to achieve this, the knowledge of a time variable was
essential.

Existence of the spinorial wavefunctions can be interpreted as an
inequivalent quantization of the theory, since there are no physical
observables which mix the two UIR. Note that Carlip has found a spinorial
wavefunction in the Chern-Simmons formulation of the theory on a torus [18].
Since the Chern-Simmons
formulation is canonically equivalent to the Witten formulation,
and $C^*_W$ is a subset of $C^*_A$, then it is not surprising that a spinorial
wavefunction appears in the Witten theory.

Generalizing our method to the higher genus case does not appear
to be straightforward, although a natural candidate for the role of
the forms $\chi$ is the cohomology basis on $\Si$, consisting
of $2g$ forms, where $g$ is the
genus of $\Si$.

\noindent{\bf Acknowledgements}

We would like to thank A. Ashtekar, K. Kucha\v r and C. Isham for the
discussions.

\noindent{\bf References}

\frenchspacing
\item{[1]}S. Deser, R. Jackiw and G. `t Hooft, Annals of Physics 152
(1984) 220
\item{[2]}A. Ashtekar, Non-Perturbative Canonical Gravity, World Scientific,
Singapore (1991)
\item{[3]}C. Rovelli and L. Smolin, Nucl. Phys. B 331 (1990) 80
\item{[4]}E. Witten, Nucl. Phys. B 311 (1988) 46
\item{[5]} I. Bengtsson, Phys. Lett. 220B (1989) 51
\item{[6]} J.E. Nelson and T. Regge, Nucl. Phys. B328 (1989) 190
\item{[7]} S. Carlip, Nucl. Phys. B324 (1989) 106
\item{[8]}A. Ashtekar, V. Husain, C. Rovelli, J. Samuel, L. Smolin, Class.
Quant. Grav.  6 (1989) L 185
\item{[9]} H. Kodama, Prog. Theor. Phys. 80 (1988) 1024
\item{[10]} N. Manojlovi\'c and A. Mikovi\'c, Canonical Analysis of the
Bianchi Models in the Ashtekar Formulation, Imperial and QMW preprint,
Imperial/TP/90-91/27 and QMW/PH/91/12 (1991)
\item{[11]} A. Bracken and B. Jessup J. Math. Phys 23 (1982) 1925
\item{[12]} W. Siegel, Introduction to String Field Theory, World Scientific,
Singapore (1988)
\item{[13]} B. Binegar J. Math. Phys. 23 (1982) 1511
\item{[14]} S. Deser and  R. Jackiw Phys. Lett. B 263 (1991) 431
\item{[15]} K. Kuchar, in Quantum Gravity 2: A Second Oxford Symposium, eds.
C.J. Isham, R. Penrose and D.W. Sciama, Clarendon Press, Oxford (1981)
\item{[16]} S.S. Schweber, An Introduction to Relativistic Field Theory,
Peterson and Row, Evanston (1961)
\item{[17]} C. Rovelli, Ashtekar Formulation of General Relativity and Loop
Space Non-Perturbative Quantum Gravity, University of Pittsburgh preprint
(1991)
\item{[18]} S. Carlip, (2+1)-Dimensional Chern-Simons Gravity as a Dirac Square
Root, UC Davis preprint, UCD-91-16 (1991)
\bye